# Tunneling valley Hall effect driven by tilted Dirac fermions

Shu-Hui Zhang[1,*] Ding-Fu Shao[2,†] Zi-An Wang,[2,3] Jin Yang,[4] Wen Yang[4,‡] and Evgeny Y. Tsymbal[5,§]

[1] *College of Mathematics and Physics, Beijing University of Chemical Technology, Beijing 100029, People's Republic of China*

[2] *Key Laboratory of Materials Physics, Institute of Solid State Physics, HFIPS, Chinese Academy of Sciences, Hefei 230031, China*

[3] *University of Science and Technology of China, Hefei 230026, China*

[4] *Beijing Computational Science Research Center, Beijing 100193, China*

[5] *Department of Physics and Astronomy & Nebraska Center for Materials and Nanoscience, University of Nebraska, Lincoln, Nebraska 68588-0299, USA*

Valleytronics is a research field utilizing a valley degree of freedom of electrons for information processing and storage. A strong valley polarization is critical for realistic valleytronic applications. Here, we predict a tunneling valley Hall effect (TVHE) driven by tilted Dirac fermions in all-in-one tunnel junctions based on a two-dimensional (2D) valley material. Different doping of the electrode and spacer regions in these tunnel junctions results in momentum filtering of the tunneling Dirac fermions, generating a strong transverse valley Hall current dependent on the Dirac-cone tilting. Using the parameters of an existing 2D valley material, we demonstrate that such a TVHE is much stronger than that induced by the intrinsic Berry curvature mechanism reported previously. Finally, we predict that resonant tunneling can occur in a tunnel junction with properly engineered device parameters such as the spacer width and transport direction, providing significant enhancement of the valley Hall angle. Our work opens a new approach to generate valley polarization in realistic valleytronic systems.

Electrons in solids possess internal degrees of freedom, such as charge, spin, and orbital, which can be employed as state variables to encode and process information. Recently, valley, *i.e.*, a local well-separated extremum of an energy band in the momentum space, has been identified as an additional degree of freedom carried by low-energy electrons [1,2]. The exploitation and manipulation of the valley degree of freedom manifests an emerging field of research known as valleytronics [3-20]. To date, many material systems have been discovered exhibiting valleys in their band structure [21-23]. For example, in two-dimensional (2D) materials, such as graphene and transition metal dichalcogenides, there are two inequivalent valleys that occur at the $K$ and $\bar{K}$ ($-K$) points at the edges of the Brillouin zone. In a doped system, an imbalance in carrier population between the $K$ and $\bar{K}$ valleys, known as valley polarization, could be used to store binary information. However, the valleys at the reversed momenta, $K$ and $\bar{K}$, in these materials are usually degenerate, which makes their polarization elusive.

Significant efforts have been made to break the valley degeneracy and generate the valley polarization. Since $K$ and $\bar{K}$ are connected by time reversal symmetry ($\hat{T}$), the perturbations breaking $\hat{T}$ symmetry, such as optical [6, 24-33] and magnetic [29, 34 - 44], have been employed. These effects, though interesting, require large perturbative fields inaccessible for practical use. It was suggested that the large valley polarization can be induced in magnetic materials or heterostructures, where the spin and valley degrees of freedom are strongly coupled [45, 46]. This mechanism, however, does not work for $\hat{T}$ invariant nonmagnetic materials.

Among the existing mechanisms, a valley Hall effect (VHE) allows the electrical generation of transverse valley dependent Hall currents by applying a longitudinal valley-neutral charge current, resulting in the accumulation of $K$ and $\bar{K}$ valley carriers on different edges [2]. This produces a detectable valley polarization in real space, while not breaking the valley degeneracy in $k$-space. The intrinsic VHE is driven by the opposite-sign Berry curvatures at $K$ and $\bar{K}$, leading to the opposite transverse velocities at the two valleys [2]. However, the Berry curvature only emerges in systems where the Kramers degeneracy is lifted, i.e., in nonmagnetic materials with strong spin-orbit coupling, where space inversion ($\hat{P}$) symmetry is broken by a non-centrosymmetric structure or by the application of an electric field [47], or in magnetic materials, where $\hat{T}$ symmetry is broken by magnetism. It is sizable only in the semiconductors with very small band gaps. This limits the material choice of VHE and also demands a very strict control of the stoichiometry, as the doping effect can easily shift the Fermi energy ($E_F$) away from the band gap and significantly reduce the Berry curvature. Therefore, although VHE is promising for all-electric valleytronics, the output signal is usually too weak for realistic applications. Thus, it would be interesting from the fundamental point of view and desirable for valleytronic applications to find a new mechanism of VHE independent of the Berry curvature.

Here, we predict a Berry-curvature-free tunneling VHE (TVHE) produced by tilted Dirac fermions. The effect occurs in an all-in-one tunnel junction based on a 2D valley material where the valleys are described by the tilted Dirac Hamiltonian, and the electrodes and the spacer layer have different doping. We find that the tunneling Dirac fermions carry net transverse velocities opposite for $K$- and $\bar{K}$- valleys due to the momentum filtering in the spacer layer controlled by the Dirac-cone tilting. This generates a strong TVHE with a giant valley Hall angle that is defined as the ratio of the valley Hall conductance and the longitudinal conductance. Finally, we predict that resonant tunneling can occur in a tunnel junction with properly engineered device parameters such as spacer width and transport direction, providing significant enhancement of the valley Hall angle.

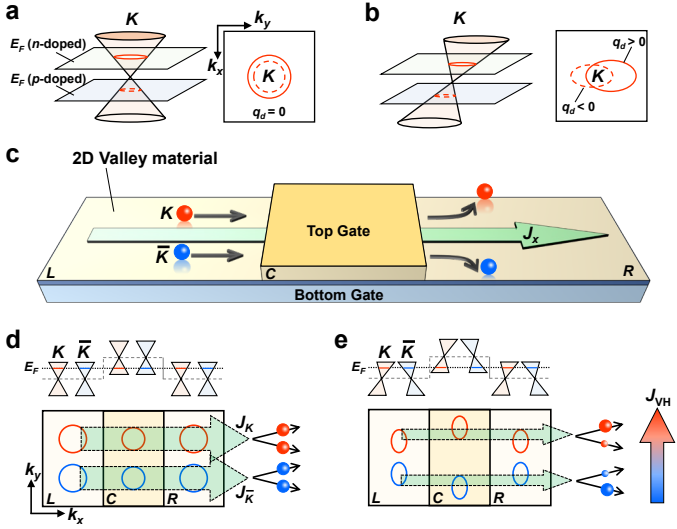

**FIG. 1:** (**a,b**) Schematic electronic structure of a 2D material with a valley arising an untitled (a, left) and titled (b, left) Dirac cone at the $K$ point with respective Fermi surfaces of untitled (a, right) and tilted (b, right) Dirac fermions upon doping. (**c**) A schematic of all-in-one tunnel junction representing left (L) and right (R) electrode regions separated by central (C) spacer region. Top and bottom gates control doping in the electrodes and the spacer. A valley-neutral current $J_x$ generates TVHE in the right electrode. (**d,e**) The origin of the TVHE. For untitled Dirac cone valleys, the transverse electron velocities at $K$ and $\bar{K}$ are perfectly balanced, prohibiting the TVHE (d). For tilted Dirac cone valleys, electrons with negative (positive) transverse velocities at $K$ ($\bar{K}$) are largely filtered out, leading to a sizable TVHE and the associated transverse valley Hall current $J_{VH}$ in the right electrode.

We consider a 2D material with valleys originating from an untitled (Fig 1(a)) or titled (Fig 1(b)) Dirac cones at momenta $K$ and $\bar{K}$ (Fig. 1(a,b)). We assume an all-in-one tunnel junction based on this 2D valley material, which is divided into the left (L) and right (R) electrode regions and the central (C) spacer region. The electrodes are assumed to be $n$-doped and the spacer $p$-doped by voltage applied from the top and bottom gates. The valleys at the $K$ and $\bar{K}$ points in this 2D material are described by the Dirac Hamiltonian [48]

$$H_{\eta,i}(\boldsymbol{q}) = \eta\mu_1\sigma_1 q_X + \mu_2\sigma_2 q_Y + \eta\mu_t\sigma_0 q_Y + V_i. \quad (1)$$

Here, index $i$ denotes the left ($i = $ L) or right ($i = $ R) electrodes or the spacer ($i = $ C). The Cartesian coordinate system $X$-$Y$ is used, where the $X$ ($Y$) axis is perpendicular (parallel) to the tilt direction, $\eta = 1\,(-1)$ denotes the $K$- ($\bar{K}$-) valleys, $\boldsymbol{q} = (q_X, q_Y)$ is the wave vector with respect to $K$ or $\bar{K}$, $\sigma_1$ and $\sigma_2$ are the Pauli matrices, $\sigma_0$ is the identity matrix, parameters $\mu_1$, $\mu_2$, and $\mu_t$ control the anisotropy and the tilt of the Dirac cone, and $V_i$ are the potential associated with the doping effect.

In the absence of gate voltage ($V_i = 0$), the Dirac points are located at the Fermi level, $E_F = 0$, for all regions in the junction. In the presence of gate voltage, the Dirac point in region $i$ is shifted to $E_F + V_i$ due to the doping effect produced by the gate voltage. When $\mu_1 = \mu_2$ and $\mu_t = 0$, the Dirac cone is not tilted, resulting in the isotropic Dirac fermions, as in graphene [49]. The associated Fermi surface is a circle centered at $q_X = q_Y = 0$. It changes size but does not shift upon doping (Fig. 1(a)). When $\mu_t \neq 0$, the Dirac cone is tilted along the $Y$-direction, resulting in type-I tilted Dirac fermions. The associated Fermi surface is an ellipse centered at $q = (0, \eta q_d)$. It changes size and shifts along the $Y$-direction upon doping, with the shift $q_d$ having opposite sign for $n$- and $p$-doing (Fig. 1(b)). The type-I tilted Dirac fermions have been observed in many 2D materials such as strained graphene [48], α-(BEDT-TTF)$_2$I$_3$ [50], and 8-Pmmn borophene [51,52]. There can be also type-II Dirac fermions induced by strong tilting (e.g., [53]), which are not considered in this work since they cannot represent valleys as the Dirac points are not at the band extrema.

The longitudinal conductance is given by [54]

$$\sigma_{xx} = \frac{\sigma_0}{(2\pi)^2} \int \left( T_{q_y}^K + T_{q_y}^{\bar{K}} \right) dq_y, \quad (2)$$

where $x$ ($y$) is the longitudinal (transverse) transport direction, $T_{q_y}^\eta$ is transmission for valley $\eta$ at the transverse wave vector $k_y$, and $\sigma_0 = 2e^2/\hbar$. Hall conductance $\sigma_{yx}$ and valley Hall conductance $\sigma_{yx}^V$ in the right electrode are given by [55-60]

$$\sigma_{yx} = \sigma_{yx}^K + \sigma_{yx}^{\bar{K}}$$
$$\sigma_{yx}^V = \sigma_{yx}^K - \sigma_{yx}^{\bar{K}} \quad (3)$$

where $\sigma_{yx}^\eta$ is the conductance associated with the valley-dependent transverse current $J_y^\eta$:

$$\sigma_{yx}^\eta = \frac{\sigma_0}{(2\pi)^2} \int \zeta_{q_y}^{\eta,\rightarrow} T_{q_y}^\eta dq_y, \quad (4)$$

where $\zeta_{q_y}^{\eta,\rightarrow} = \frac{v_{y,q_y}^{\eta,\rightarrow}}{v_{x,q_y}^{\eta,\rightarrow}}$. Here $v_{r,q_y}^{\eta,\rightarrow}$ denotes the longitudinal ($r = x$) and transverse ($r = y$) band velocities of the right-going ($\rightarrow$) states in the electrode. The Hall angle $\Theta = \sigma_{yx}/\sigma_{xx}$ and valley Hall angle $\Theta_V = \sigma_{yx}^V/\sigma_{xx}$ can be then defined to estimate the strengths of these effects. We set $E_F = 0$ and assume that the left and the right electrodes are $n$-doped with $V_L = V_R = -V_0$, and the central region is $p$-doped with $V_C = V_0$. In the ballistic transport regime with negligible intervalley scattering and conserved wave vector, and $T^\eta_{q_y}$ is determined by matching the Fermi surfaces in the electrodes and the spacer. When the tilt direction $Y$ is perpendicular to the longitudinal transport direction (Fig. 1(b)), the Fermi surface of valley $\eta$ is symmetric with respect to $q_y = \eta q_d$, resulting in $v^{\eta,\to}_{x,q_y} = v^{\eta,\to}_{x,q'_y}$ and $v^{\eta,\to}_{y,q_y} = -v^{\eta,\to}_{y,q'_y}$, where $q'_y = 2\eta q_d - q_y$. This leads to $\zeta^{\eta,\to}_{q_y} = -\zeta^{\eta,\to}_{q'_y}$. Therefore, the distribution of $T^\eta_{q_y}$ determines $\sigma_{yx}$.

In the absence of tilting, the valley Fermi surfaces are perfectly circular with different sizes in the electrode and spacer regions due to different doping, as shown in Fig. 1(a,d). Since the origin of these Fermi surfaces does not change with doping ($q_d = 0$), the tunneling barrier height for Dirac fermions is momentum- and valley- independent, resulting in $T^K_{q_y} = T^K_{-q_y} = T^{\bar{K}}_{q_y} = T^{\bar{K}}_{-q_y}$. Therefore, the transmitted Dirac fermions have zero net transverse velocities, for both total current $J_x$ and valley-dependent current $J^\eta_x$, resulting in zero Hall and valley Hall conductances (Fig. 1(d)).

In contrast, in the presence of tilting, the Fermi surfaces of Dirac fermions are shifted and elongated along the $y$ ($Y$) direction, as shown in Fig. 1(b,e). Since the displacement of the Fermi surfaces in the electrodes and the spacer are opposite in sign due to opposite ($n$- and $p$-) doping, the tunneling barrier height is momentum-dependent, resulting in *momentum filtering* that creates disbalance between the tunneling Dirac fermions at $q_y$ and $q'_y$ at each valley, leading to $T^\eta_{q_y} \neq T^\eta_{q'_y}$ and hence the finite valley-dependent transverse current $J^\eta_y$. However, for tilting along the $y$ ($Y$) direction, each valley is symmetric with respect to $q_x = 0$ and hence $v^{\eta,\to}_{x,q_y} = -v^{\eta,\leftarrow}_{x,q_y}$ and $v^{\eta,\to}_{y,q_y} = v^{\eta,\leftarrow}_{y,q_y}$, where $\leftarrow$ denotes the left-going state. In addition, the two valleys are connected by $\hat{T}$ symmetry so that $v^{K,\to}_{x,q_y} = -v^{\bar{K},\leftarrow}_{x,-q_y}$, $v^{K,\to}_{y,q_y} = -v^{\bar{K},\leftarrow}_{y,-q_y}$ and hence $\zeta^{K,\to}_{q_y} = -\zeta^{\bar{K},\to}_{-q_y}$. $\hat{T}$ symmetry also enforces $T^K_{q_y} = T^{\bar{K}}_{-q_y}$ and thus $\sigma^K_{yx} = -\sigma^{\bar{K}}_{yx}$, resulting in the valley-dependent transverse currents opposite in sign, $J^K_y = -J^{\bar{K}}_y$. Therefore, a non-zero valley Hall current $J_{VH} = J^K_y - J^{\bar{K}}_y = 2J^K_y$ is expected with the valley Hall conductance given by

$$\sigma^V_{yx} = \frac{2\sigma_0}{(2\pi)^2} \int \zeta^{K,\to}_{q_y} T^K_{q_y} dq_y. \quad (5)$$

Due to $\zeta^{\eta,\to}_{q_y} = -\zeta^{\eta,\to}_{q'_y}$, a non-zero $\sigma^V_{yx}$ occurs if $T^K_{q_y} \neq T^K_{q'_y}$ which is expected for a tilted Dirac cone.

Next, we quantitatively demonstrate TVHE by performing quantum-transport calculations for a realistic tunnel junction.

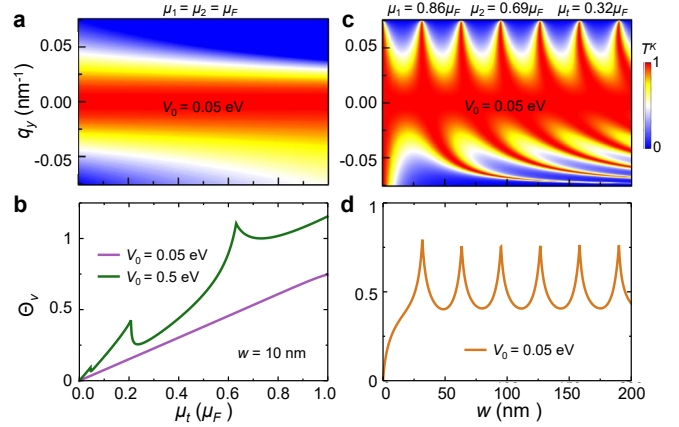

**FIG. 2:** (**a,b**) Transmission $T^K_{q_y}$ for valley $K$ as a function of $q_y$ (a) and valley Hall angle $\Theta_V$ as a function $\mu_t$ (b) for $w = 10$ nm. (**c,d**) Transmission $T^K_{q_y}$ (c) and $\Theta_V$ valley Hall angle (d) as functions of $w$. The parameters used in the calculations are indicated.

We first consider $\mu_1 = \mu_2 = \mu_F = 10^6$ m/s and $\mu_t = 0$, corresponding to isotropic Dirac fermions in graphene [49], and assume central region width $w = 10$ nm and low doping of $V_0 = 0.05$ eV. As expected, we find that $T^K_{q_y}$ is symmetric with respect to $q_y = 0$ (Fig. 2(a) for $\mu_t = 0$), leading to vanishing $\sigma_{yx}$ and $\sigma^V_{yx}$. For non-zero $\mu_t$, transmission $T^K_{q_y}$ becomes asymmetric with respect to $q_y$ (Fig. 2(a) for $\mu_t > 0$), which implies $T^K_{q_y} \neq T^K_{q'_y}$ giving rise to a finite $\sigma^V_{yx}$ (Fig. 2(b)). For a weak tilting of $\mu_t = 0.2\ \mu_F$, we obtain a sizable valley Hall angle $\Theta_V = 0.15$. $\Theta_V$ can be further enhanced by increasing $\mu_t$ (Fig. 2(b)).

For high doping of $V_0 = 0.5$ eV, we find a much larger $\Theta_V$. This is expected as the increase of $V_0$ shifts the Fermi surfaces stronger upon doping and thus enhances momentum filtering asymmetry between at $q_y$ and $q'_y$ at each valley and thus $J^K_y$ and $J_{VH} = 2J^K_y$ (Fig. 1(e)). Interestingly, we find that the $\Theta_V$-versus-$\mu_t$ curve exhibits anomalous peaks (Fig. 2(b)). Such peaks originate from resonant tunneling due to the matching of the spacer width $w$ and the longitudinal electronic wavelength $\lambda_x$ in the spacer region, such that $w = 0.5n\lambda_x$, where $n$ is an integer, $\lambda_x = 2\pi/q_x$, and $q_x = \frac{1}{\mu_1}\sqrt{(E_F - V_C - \mu_t q_y)^2 - \mu_2^2 q_y^2}$ [61].

The emergence of resonant tunneling implies a possibility to realize a strong VHE with a low doping and moderate tilting. Here, we consider parameters $\mu_1 = 0.86\ \mu_F$, $\mu_2 = 0.69\ \mu_F$ and $\mu_t = 0.32\ \mu_F$ corresponding to 2D valley material 8-*Pmmn* borophene [52], and calculate $T^K_{q_y}$ and $\Theta_V$ as functions of $w$. For a low doping of $V_0 = 0.05$ eV, we find that the distribution of $T^K_{q_y}$ exhibits a "fishbone"-like pattern (Fig. 2(c)), which has typical peaks reflecting the resonant tunneling. For example, for $q_y = 0.05$ nm$^{-1}$, the resonant enhancements occur at $w = 0.5n\lambda_x \approx 32n$ nm, as can be seen from Fig. 2(c). As a result, a large $\Theta_V \approx 0.75$ is predicted for certain widths of the spacer layer $w$ (Fig. 2(d)).

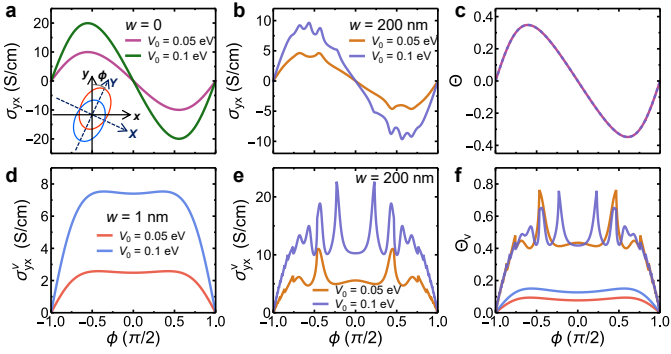

**FIG. 3:** (a,b) Hall conductance $\sigma_{yx}$ as a function of angle $\phi$ for $w = 0$ (a) and 200 nm (b) for $V_0 = 0.05$ and 0.1 eV. (c) Hall angle $\Theta_V$ as a function of $\phi$ for the parameters as in (a) and (b) (d,e) Valley Hall conductance $\sigma_{yx}^V$ as a function of $\phi$ for $w = 1$ nm (d) and 200 nm (e). (f) Valley Hall angle $\Theta_V$ as a function of $\phi$ for the parameters as in (d) and (e).

We further consider a more general case where the tilting $Y$-direction has an angle $\phi$ away from the transverse $y$-direction (inset in Fig. 3(a)), using the parameters for 8-$Pmmn$ borophene [52]. Experimentally, the angle $\phi$ can be controlled by changing the direction of the longitudinal current. We choose the central region widths $w$ such that they do not support resonant tunneling and calculate Hall conductance $\sigma_{yx}$ as a function of $\phi$ for different $V_0$. For $\phi \neq 0.5\,n\pi$ ($n$ is an integer), we find that the symmetry constraint $\sigma_{yx}^K = -\sigma_{yx}^{\bar{K}}$ is absent, resulting in a finite $\sigma_{yx}$ even for a uniform system ($w = 0$, Fig. 3(a)). This is due to the fact that at $\mu_1 \neq \mu_2$, $\sigma_{xx}(\phi = 0) \neq \sigma_{xx}(\phi = 0.5\pi)$ resulting in $\sigma_{yx}(\phi) = [\sigma_{xx}(0.5\pi) - \sigma_{xx}(0)]\sin\phi\cos\phi$. When $w$ is large, resonant tunneling appears and leads to peaks in $\sigma_{yx}$ (Fig. 3(b)) Interestingly, although $w$ and $V_0$ strongly influence the $\sigma_{yx}$-$\phi$ curve, the calculated Hall angles $\Theta$ are the same for different $w$ and $V_0$, and the features of resonant tunneling are not revealed in $\Theta$ (Fig. 3(c)). On the contrary, the valley Hall conductance $\sigma_{yx}^V$ arises only at nonzero $w$ and $\phi \neq 0.5\,m\pi$ ($m$ is an odd number) (Fig. 3(d,e)) and increases with increasing $V_0$. For a small $w$, the $\sigma_{yx}^V$-$\phi$ curves are smooth (Fig. 3(d)), while for a large $w$, pronounced peaks associated with resonant tunneling emerge (Fig. 3(e)). These peaks also appear in the $\Theta_V$-$\phi$ curves (Fig. 3(f)). Therefore, transport direction can also be used to control resonant tunneling for enhancing $\Theta_V$.

The different behaviors of $\Theta$ and $\Theta_V$ shown in Figs. 3(c) and 3(f) can be understood from a simple analytic derivation. Since the two valleys are connected by $\hat{T}$ symmetry, we have $\mathcal{T}_{q_y} \equiv T_{q_y}^K = T_{-q_y}^{\bar{K}}$ and $v_{r,q_y}^{K,\rightarrow} = -v_{r,-q_y}^{\bar{K},\leftarrow}$ (where $\leftarrow$ denotes the left-going state). The latter results in $\zeta_{q_y}^{K,\rightarrow} = \zeta_{-q_y}^{\bar{K},\leftarrow}$. We can then define $\vartheta_\pm = \zeta_{q_y}^{K,\rightarrow} \pm \zeta_{q_y}^{K,\leftarrow}$ and rewrite Eq. (3) as follows

$$\sigma_{yx} = \frac{\sigma_0}{(2\pi)^2}\int \mathcal{T}_{q_y}\vartheta_+\,dq_y,$$
$$\sigma_{yx}^V = \frac{\sigma_0}{(2\pi)^2}\int \mathcal{T}_{q_y}\vartheta_-\,dq_y. \quad (6)$$

Introducing parameters $\gamma_1 = \mu_2/\mu_1$ and $\gamma_2 = \mu_t/\mu_1$ to reflect the anisotropy and tilting, respectively, we obtain [62]

$$\vartheta_+ = \frac{2\sin\phi\cos\phi\,(\gamma_1^2 - \gamma_2^2 - 1)}{\cos^2\phi + (\gamma_1^2 - \gamma_2^2)\sin^2\phi},$$
$$\vartheta_- = \frac{(\gamma_2 V_0 \mu_1 \cos\phi + (\gamma_1^2 - \gamma_2^2)q_y\mu_1^2)(q_x^\leftarrow - q_x^\rightarrow)}{(\gamma_1^2 - \gamma_2^2)q_y^2\mu_1^2 + 2\gamma_2 V_0\mu_1(\cos\phi + 1) - (\cos^2\phi + \gamma_1^2\sin^2\phi)V_0^2}, \quad (7)$$

It is seen that $\vartheta_+$ is a function of $\gamma_1, \gamma_2, \phi$ and does not depend on $w$, $V_0$ and $q_y$. As a result, the Hall angle takes a universal form

$$\Theta \equiv \frac{\sigma_{yx}}{\sigma_{xx}} = \frac{\sin\phi\cos\phi\,(\gamma_1^2 - \gamma_2^2 - 1)}{\cos^2\phi + (\gamma_1^2 - \gamma_2^2)\sin^2\phi}, \quad (8)$$

consistent with our numerical result in Fig. 2(c). On the other hand, $\vartheta_-$ is dependent on $V_0$ and $q_y$. This leads to $\Theta_V$ being dependent on $w$ and $V_0$, as shown in Fig. 3(f).

The predicted TVHE is related to the previously proposed tunneling anomalous and spin Hall effects [55-60]. Both effects are controlled by the momentum-dependent tunneling barrier height and the associated momentum filtering. However, the origins of the momentum filtering for these effects are different. The tunneling anomalous and spin Hall effects occur due to the symmetry mismatch in a magnetic electrode, where $\hat{T}$ symmetry is broken, and a nonmagnetic barrier layer with strong spin-orbit coupling, where $\hat{T}$ symmetry is preserved [55-60]. On the contrary, for the TVHE, the momentum-filtering is due to the mismatch of the Fermi surfaces of the tilted Dirac fermions in the electrode and spacer regions due to opposite doping. The symmetry remains the same for all regions of such a tunnel junction, and hence the junction can be constructed using a single 2D material (Fig. 1), where the doping, the spacer width, and the transport direction can be conveniently controlled.

The proposed TVHE has significant advantages over the intrinsic VHE driven by the Berry curvature mechanism. First, the intrinsic VHE occurs only in materials with finite spin-orbit coupling and broken $\hat{P}\hat{T}$ symmetry. The proposed TVHE, on the other hand, does not have these restrictions, and thus can emerge even in nonmagnetic and centrosymmetric materials composed of light elements. Second, unlike the intrinsic VHE where a valley Hall angle is usually very small [63], the proposed TVHE exhibits a giant valley Hall angle that can be further optimized by engineering parameters of the junction such as Dirac cone tilting, doping, spacer width, and transport direction.

The TVHE is expected to occur not only for the oppositely doped electrodes and spacer ($V_L = V_R = -V_C$), as was assumed above, but for any modulated doping along the junction. As long as $V_L = V_R \neq V_C$, a nonvanishing transverse valley current emerges and produces the TVHE. In addition to 2D materials with massless tilted Dirac fermions considered in this work, such as strained graphene [48], α-(BEDT-TTF)$_2$I$_3$ [50] and 8-Pmmn borophene [51,52], there exist 2D materials hosting massive tilted Dirac fermions, such as monolayer 1$T'$-MX$_2$ (M = Mo, W, X=S, Se, Te) [64], which can be also used to realize the proposed TVHE.

Finally, the proposed TVHE device can be made non-volatile. In a heterostructure where the 2D valley material is placed on top of a ferroelectric insulator, different doping of the electrodes and the spacer layer can be induced by polarization charges of oppositely polarized ferroelectric domains, and the TVHE can be controlled by ferroelectric switching. This opens a possibility for non-volatile valleytronic applications.

In conclusion, we have predicted a TVHE driven by the tilted Dirac fermions in an all-in-one tunnel junction based on a two-dimensional (2D) valley material. Different doping of the electrodes and the spacer layer in this tunnel junction results in tilting-dependent momentum filtering of the tunneling Dirac fermions, generating a strong TVHE with a giant valley Hall angle. Resonant tunneling is predicted to occur in the tunnel junction with properly engineered device parameters, such as spacer width and transport direction, providing significant enhancement of the valley Hall angle. Our work opens a new approach to generate valley polarization in realistic valleytronic systems.

**Acknowledgments.** This work was supported by the National Key R&D Program of China (Grant Nos. 2017YFA0303400, 2021YFA1600200), the National Natural Science Foundation of China (NSFC; Grants Nos. 12174019, 11774021, 12274411, 12241405, 52250418), and the NSAF grant in NSFC (Grant No. U1930402). E.Y.T. acknowledges support from the EPSCoR RII Track-1 (NSF Award OIA-2044049) program. The authors acknowledge the computational support from the Beijing Computational Science Research Center (CSRC) and Hefei Advanced Computing Center. The figures were created using the SciDraw scientific figure preparation system [65].

* shuhuizhang@mail.buct.edu.cn

† dfshao@issp.ac.cn

‡ wenyang@csrc.ac.cn

§ tsymbal@unl.edu